\documentclass[prl,twocolumn,superscriptaddress]{revtex4-2}
\usepackage{dcolumn}
\usepackage{mathtools}
\usepackage{amssymb}
\usepackage{bm}
\usepackage{pifont}
\usepackage{psfrag}
\usepackage{epstopdf}
\usepackage{wasysym}
\usepackage{amsmath}
\usepackage{tipa}
\usepackage{hyperref}
\usepackage{setspace}
\usepackage[usenames]{color}
\emergencystretch=\maxdimen
\hypersetup{
     colorlinks = true,
     linkcolor = blue,
     anchorcolor = blue,
     citecolor = blue,
     filecolor = blue,
     urlcolor = blue
     }
\renewcommand{\footnoterule}{  \hrule width \textwidth height 1pt
  \kern 2pt}

\begin{document}

\title{Enhancement of quantum heat engine by encircling a Liouvillian exceptional point}
\author{J.-T.~Bu}  \thanks{Co-first authors with equal contribution}
\affiliation{State Key Laboratory of Magnetic Resonance and Atomic and Molecular Physics, Wuhan Institute of Physics and Mathematics, Innovation Academy of Precision Measurement Science and Technology, Chinese Academy of Sciences, Wuhan 430071, China}
\affiliation{University of the Chinese Academy of Sciences, Beijing 100049, China}
\author{J.-Q.~Zhang} \thanks{Co-first authors with equal contribution}
\affiliation{State Key Laboratory of Magnetic Resonance and Atomic and Molecular Physics, Wuhan Institute of Physics and Mathematics, Innovation Academy of Precision Measurement Science and Technology, Chinese Academy of Sciences, Wuhan 430071, China}
\author{G.-Y.~Ding}  \thanks{Co-first authors with equal contribution}
\affiliation{State Key Laboratory of Magnetic Resonance and Atomic and Molecular Physics, Wuhan Institute of Physics and Mathematics, Innovation Academy of Precision Measurement Science and Technology, Chinese Academy of Sciences, Wuhan 430071, China}
\affiliation{University of the Chinese Academy of Sciences, Beijing 100049, China}
\author{J.-C.~Li}
\affiliation{State Key Laboratory of Magnetic Resonance and Atomic and Molecular Physics, Wuhan Institute of Physics and Mathematics, Innovation Academy of Precision Measurement Science and Technology, Chinese Academy of Sciences, Wuhan 430071, China}
\affiliation{University of the Chinese Academy of Sciences, Beijing 100049, China}
\author{J.-W.~Zhang}
\affiliation{Research Center for Quantum Precision Measurement, Guangzhou Institute of Industry Technology, Guangzhou, 511458, China }
\author{B.~Wang}
\affiliation{State Key Laboratory of Magnetic Resonance and Atomic and Molecular Physics, Wuhan Institute of Physics and Mathematics, Innovation Academy of Precision Measurement Science and Technology, Chinese Academy of Sciences, Wuhan 430071, China}
\affiliation{University of the Chinese Academy of Sciences, Beijing 100049, China}
\author{W.-Q.~Ding}
\affiliation{State Key Laboratory of Magnetic Resonance and Atomic and Molecular Physics, Wuhan Institute of Physics and Mathematics, Innovation Academy of Precision Measurement Science and Technology, Chinese Academy of Sciences, Wuhan 430071, China}
\affiliation{University of the Chinese Academy of Sciences, Beijing 100049, China}
\author{W.-F.~Yuan}
\affiliation{State Key Laboratory of Magnetic Resonance and Atomic and Molecular Physics, Wuhan Institute of Physics and Mathematics, Innovation Academy of Precision Measurement Science and Technology, Chinese Academy of Sciences, Wuhan 430071, China}
\affiliation{University of the Chinese Academy of Sciences, Beijing 100049, China}
\author{L.~Chen}
\affiliation{State Key Laboratory of Magnetic Resonance and Atomic and Molecular Physics, Wuhan Institute of Physics and Mathematics, Innovation Academy of Precision Measurement Science and Technology, Chinese Academy of Sciences, Wuhan 430071, China}
\affiliation{Research Center for Quantum Precision Measurement, Guangzhou Institute of Industry Technology, Guangzhou, 511458, China }
\author{S.~K.~\"{O}zdemir}
\email{sko9@psu.edu}
\affiliation{Department of Engineering Science and Mechanics, and Materials Research Institute, Pennsylvania State University, University Park, State College, Pennsylvania 16802, USA }
\author{F.~Zhou}
\email{zhoufei@wipm.ac.cn}
\affiliation{State Key Laboratory of Magnetic Resonance and Atomic and Molecular Physics, Wuhan Institute of Physics and Mathematics, Innovation Academy of Precision Measurement Science and Technology, Chinese Academy of Sciences, Wuhan 430071, China}
\affiliation{Research Center for Quantum Precision Measurement, Guangzhou Institute of Industry Technology, Guangzhou, 511458, China }
\author{H.~Jing}
\email{jinghui73@foxmail.com}
\affiliation{Key Laboratory of Low-Dimensional Quantum Structures and Quantum Control of Ministry of Education, Department of Physics and Synergetic Innovation Center for Quantum Effects and Applications, Hunan Normal University, Changsha 410081, China}
\author{M.~Feng}
\email{mangfeng@wipm.ac.cn}
\affiliation{State Key Laboratory of Magnetic Resonance and Atomic and Molecular Physics, Wuhan Institute of Physics and Mathematics, Innovation Academy of Precision Measurement Science and Technology, Chinese Academy of Sciences, Wuhan 430071, China}
\affiliation{Research Center for Quantum Precision Measurement, Guangzhou Institute of Industry Technology, Guangzhou, 511458, China }
\affiliation{School of Physics, Zhengzhou University, Zhengzhou 450001, China}
%\\
%\affiliation{$^{1}$ State Key Laboratory of Magnetic Resonance and Atomic and Molecular Physics, Wuhan Institute of Physics and Mathematics,
%Innovation Academy of Precision Measurement Science and Technology, Chinese Academy of Sciences, Wuhan, 430071, China\\
%$^{2}$ School of Physics, University of the Chinese Academy of Sciences, Beijing 100049, China\\
%$^{3}$ Research Center for Quantum Precision Measurement, Guangzhou Institute of Industry Technology, Guangzhou, 511458, China, \\
%$^{4}$ \\
%$^{5}$ \\
%$^{6}$ School of Physics, Zhengzhou University, Zhengzhou 450001, China }

\begin{abstract}
Quantum heat engines are expected to outperform the classical counterparts due to quantum coherences involved.  Here we experimentally execute a single-ion quantum heat engine and demonstrate, for the first time, the dynamics and the enhanced performance of the heat engine originating from the Liouvillian exceptional points (LEPs). In contrast to the topological effects solely concerned for the LEPs, here we focus on the thermodynamic effects, which can be understood by the Landau-Zener-St{\"u}ckelberg process under decoherence. We witness a positive net work from the quantum heat engine if the heat engine cycle dynamically encircles an LEP. Further investigation reveals that, under the same circumstance of the heat engine cycles, the largest net work occurs when the system is operated close to the LEP. We argue that
our observation of the enhanced performance of the  quantum heat engine benefits from the eigen-energy landscape close to the LEP and the topological transition induced by the LEP. Therefore, our results open a new possibility to explore the advantages of the LEPs in quantum systems from both the thermodynamic and topological aspects.
\end{abstract}

\maketitle

Quantum heat engines (QHEs), working with quantum substances, are expected to surpass the output power and efficiency of the equivalent classical counterparts by taking advantage of quantum features~\cite{Quantumthermodynamics,Science-299-862,parado,PRL-122-110601,pra-96-052119,pra-98-042102},
such as, quantum coherence, squeezing and/or quantum correlations. The growing interest in QHEs is also fueled by the need to understand non-equilibrium thermodynamics at the nanoscale as well as the quantum-classical transition in energy-information and work-heat conversion.  Efforts have been made on exploring unique characteristics of QHEs in different quantum systems \cite{pre-86-061108,PRL-122-110601,PNAS-108-15097,PRL-112-030602,PRL-112-030602,SR-5-12953,Science-352-235,Natcommum-10-202,PRL-123-080602,PRL-100-140501,PRL-123-240601,PNAS-111-13786,PRL-125-166802,PRL-97-180402,Nonew2,PRL-112-150602,PRL-114-183602}.

As open quantum systems coupled to external thermal baths, QHEs can be considered as non-Hermitian quantum systems and may exhibit exceptional point (EP) degeneracies characterized by the coalescence of two or more eigenvalues and the associated eigenvectors of their non-Hermitian Hamiltonians~\cite{JPA-41-244018,science-363-42,nmater-18-783} or  Liouvillian superoperators~\cite{pre-97-062153}. Presence of EPs in classical and quantum systems have shown to lead many interesting and counterintuitive phenomena such as asymmetric backscattering~\cite{PNAS-113-25,Natcommum-13-599}, enhanced response to perturbations~\cite{Nature-548-187,Nature-548-192,PhotonRes-8-1457,Nature-548-187,nature-576-65,nature-576-70}, and loss-induced lasing~\cite{science-346-328}. EP-related topological features such as state exchange~\cite{PRL-126-170506}, topological energy transfer~\cite{nature-537-80}, asymmetric mode switching~\cite{nature-537-76,prx-8-021066,prl-124-153903,prl-125-187403,nature-562-86}, and enhanced phase accumulation in light-matter interactions~\cite{prl-86-787,nature-526-554}. However, these studies have considered EPs of the effective Hamiltonian or equivalently the Liouvillian formalism, which involves only coherent non-unitary evolution, ignoring quantum jumps due to decoherence. 

To capture the full dynamics of quantum systems and lay the groundwork towards EP-enabled quantum applications and processes, one should resort to Liouvillian superoperators and their exceptional points- referred to as Liouvillian exceptional points (LEPs)- which involve the interplay of energy loss and decoherence \cite{LEP}. {\color{black}Here the energy loss corresponds to coherent non-unitary evolution and the decoherence results from the quantum jump~\cite{arXiv:2111.04754}.} Recently, decoherence-enhanced phenomena near LEPs have been studied theoretically and observed in experiments as an analogue of critical damping in classical harmonic oscillators \cite{LEP,new6,new7,PRX-Quantum-2-040346,arXiv:2111.04754,PRL-127-140504}. These features are expected to play a role in full dynamical control of quantum thermal machines and their approach towards the steady state without additional time-dependent external drives \cite{PRX-Quantum-2-040346}. 

{\color{black}On the other hand, a fully dynamically encircling an LEP will lead to Landau-Zener-St{\"u}ckelberg (LZS) processes \cite{LZS} with decoherence. Conventional LZS processes result from repeated sweeps through the avoided crossing. The accumulated phase during the repeated sweeps leads to a tunability of final state probability~\cite{Science-327-669,Science-310-1653,NC-4-1401} and its applications in sensitive measurements~\cite{prl-81-2538}.}

%The topological structure of eigenvalues around the LEP leads to the chiral state transfer by dynamically encircling the LEP. This corresponds to the topological operations robust against small fluctuations of the closed trajectories. However, more effects from such topological operations with LEPs have remained largely unexplored.

\begin{figure}[tbph]
\includegraphics[width=8.5 cm]{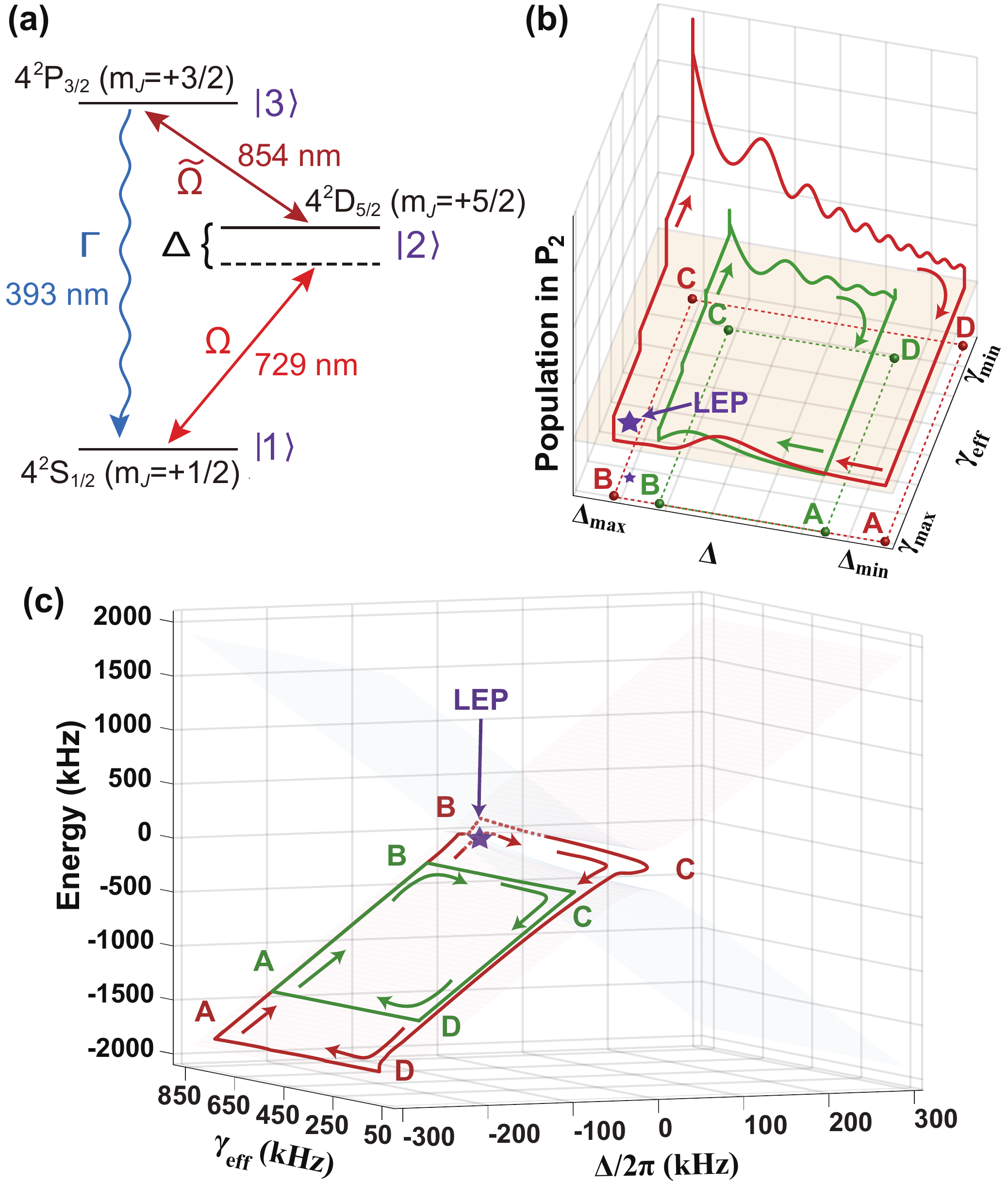}
\caption{(a) Level scheme of $^{40}$Ca$^+$ ion, where the solid arrows represent the transitions driven by lasers with Rabi frequencies $\Omega$ ($\tilde\Omega$) for the 729-nm (854-nm) laser and detuning $\Delta$ for the 729-nm laser. The wavy arrow means the spontaneous emission with decay rate $\Gamma$. (b) Simulations of the populations in $|2\rangle$ with respect to the detuning $\Delta$ and the effective decay rate $\gamma_\mathrm{eff}$, where the red (green) solid curve represents the evolution of the population with (without) the LEP encircled. The dashed curves are the projection of the solid curves with the same color on the bottom plane for guiding eyes. The four corner points A, B, C, and D are labeled for convenience of description in the text. The LEP is labeled by the purple star. (c) Illustration of trajectories with and without encircling the LEP on the Riemann surface, corresponding to the QHE cycles plotted by the solid lines with the same color in (b). In contrast to the green curve, representing the trajectory without encircling the LEP and localized just in one branch of the Riemann surface, the red curve encircling the LEP passes through the two branches (the dashed part denotes the track in the other branch), leading to the topological properties that are relevant to the thermodynamic effects observed in the present experiment. }
%Eigenenergies and (d) decay rates of the eigenvalues versus the effective decay rate $\gamma_\mathrm{eff}$ under the condition of resonance transition ($\Delta=0$). The red (green) areas represent the strong (weak) coupling regime with $\gamma_\mathrm{eff}<4\Omega$ ($\gamma_\mathrm{eff}>4\Omega$). (e) Experimental sequences for strokes of the topological heat engine.}
\label{Fig1}
\end{figure}

In this Letter, we experimentally execute a QHE using a single trapped ion by encircling an LEP and witness, for the first time, the enhanced performance of the QHE, namely, an enhanced positive net work appears when the LEP is encircled during the engine cycles. But if the engine cycle is completed without encircling the LEP, the net work can be positive or negative. In contrast to the recent concerns about the topological effects regarding the LEP, here we focus on the thermodynamic effects originating from the LEP. The enhancement in the performance of the QHE due to encircling an LEP can be understood by a Landau-Zener-St{\"u}ckelberg (LZS) process \cite{LZS} subject to decoherence. Besides, the topological transition crossing the LEP also plays an important role in this enhancement of the QHE performance.
%Since an LEP defines a topological phase transition between the exact and broken phases, the topological advantage reflected in our observation originates from the oscillatory dynamics in the exact phase. Further analysis indicates that this topological advantage also leads to maximum net work when the engine is operated close to the LEP in the exact phase.
Our experiment is carried out on a single ultracold $^{40}$Ca$^{+}$ ion confined in a linear Paul trap as employed previously \cite{SA-2-e1600578,njp-19-063032}. Under the pseudo-potential approximation, the axial and radial frequencies of the trap potential are, respectively, $\omega_z/2\pi=1.0$ MHz and $\omega_r/2\pi=1.2$ MHz. For our purpose, we employ a magnetic field of 0.6 mT directed in axial orientation, yielding the ground state $4^2S_{1/2}$ and the metastable state $3^2D_{5/2}$ split into two and six hyperfine energy levels, respectively. As shown in Fig. \ref{Fig1}(a), we encode qubit in $|4^{2}S_{1/2}, m_{J}=+1/2\rangle$ and $|3^{2}D_{5/2}, m_{J}=+5/2\rangle$, where $m_{J}$ represents the magnetic quantum number, and for simplicity the two levels are labeled as $|1\rangle$ and $|2\rangle$. To avoid thermal phonons which are detrimental to quantum effects, we perform Doppler and sideband cooling of the ion until an average phonon number of the $z$-axis motional mode is much smaller than 1 with the Lamb-Dicke parameter $\sim$0.11. The qubit is manipulated by a narrow-linewidth Ti:sapphire laser with wavelength around 729 nm, which irradiates the ultracold ion under the carrier-transition Hamiltonian $H=\Delta |2 \rangle\langle 2| + \Omega(|2 \rangle\langle 1|e^{i\phi_{L}}+|1 \rangle\langle 2|e^{-i\phi_{L}})/2$, with the detuning $\Delta$ and the Rabi frequency $\Omega$ taken in units of $\hbar=1$, as shown in Fig. \ref{Fig1}(a) and $\phi_{L}$ denoting the laser phase. In our experiment as presented below, we set $\Delta$ to vary with time and keep $\Omega$ unchanged and  $\phi_{L}=0$.

The single ultracold trapped ion is an ideal platform to explore the thermodynamics due to flexible modeling and ultimate accuracy \cite{QTE1,QTE3,QTE4,QTE5,QTE6}. Here, to implement an encircling of the LEP, we intend to manifest a two-level system with both drive and dissipation fulled engineered, and thus employ additionally the excited level $|4^{2}P_{3/2}, m_{J}=+3/2\rangle$ labeled as $|3\rangle$, with which we have a closed
cycle $|1\rangle$ $\rightarrow$ $|2\rangle$ $\rightarrow$ $|3\rangle$ $\rightarrow$ $|1\rangle$, see Fig. \ref{Fig1}(a). The first step $|1\rangle \rightarrow |2\rangle$ is achieved by a Ti:sapphire laser (729-nm) tuned exactly to the resonance transition. The second step from $|2\rangle$ to $|3\rangle$ is a dipolar transition mediated by a semiconductor laser (854-nm) under the restriction of the selection rule. The third step $|3\rangle$ $\rightarrow$ $|1\rangle$ is a spontaneous emission, also restricted by the selection rule. Practically, by tuning the 729-nm and 854-nm lasers under the condition of $\Omega\ll\tilde{\Omega}$, we may transform this three-level configuration into an effective two-level system spanned by $|1\rangle$ and $|2\rangle$ with the Rabi frequency $\Omega$ and the effective decay rate $\gamma_{\mathrm{eff}}=\tilde\Omega^{2}/\Gamma$ fully tunable \cite{PRR-2-033082}.

The dynamics of this effective two-level model is described by the Lindblad master equation,
\begin{equation}
\dot{\rho}= \mathcal{L} \rho,
\label{Eq1}
\end{equation}
where $\mathcal{L}$ is the Liouvillian superoperator, $\rho$ denotes the density operator, and $\mathcal{L}\rho\equiv -i[H,\rho]+\frac{\gamma_{\mathrm{eff}}}{2}(2|1 \rangle\langle 2|\rho|2 \rangle\langle 1|-|2 \rangle\langle 1||1 \rangle\langle 2|\rho-\rho|2 \rangle\langle 1||1 \rangle\langle 2|)$. The physics of the LEPs can be understood from the eigensolutions of $\mathcal{L}$ at $\Delta=0$, given by $\lambda_{1}=0$, $\lambda_{2}=-\gamma_{\mathrm{eff}}$, $\lambda_{3}=(-3\gamma_{\mathrm{eff}}-\xi)/4$, and $\lambda_{4}=(-3\gamma_{\mathrm{eff}}+\xi)/4$, with $\xi=\sqrt{\gamma_{\mathrm{eff}}^{2}-16\Omega^{2}}$ \cite{SM}. When $\xi=0$, that is $\gamma_{\mathrm{eff}}=4\Omega$, the eigenvalues $\lambda_{3}$ and $\lambda_{4}$ merge, giving rise to a second order LEP at $\tilde{\lambda}=-3\gamma_{\mathrm{eff}}/4$. Clearly, for $\gamma_{\mathrm{eff}}> 4\Omega$ (weak coupling), both $\lambda_{3}$ and $\lambda_{4}$ are real with a splitting amount $\xi$, corresponding to the broken phase characterized by a non-oscillatory dynamics with purely exponential decay \cite{huber,naka}. For $\gamma_{\mathrm{eff}}< 4\Omega$ (strong coupling), on the other hand, $\lambda_{3}$ and $\lambda_{4}$ form a complex conjugate pair which splits in their imaginary parts by $\xi$, corresponding to the exact phase characterized by an oscillatory dynamics. Thus, the LEP represents a topological transition point between the exact and the broken phases, dividing the parameter space into a region of oscillatory dynamics ($\gamma_{\mathrm{eff}}< 4\Omega$) and a region of non-oscillatory dynamics ($\gamma_{\mathrm{eff}}> 4\Omega$).

%the topological heat engine cycle (HEC) with LEP can ensure a positive work done. That is because topological stability can resist the suppression of population from quantum coherence. On the contrary, the one without the LEP would not. Further, with the increase of the maximum detuning, we find the maximum efficiency only can be achieved with topological HEC encircling the LEP. It results from the topological operations with the LEP containing the resonance transition in the strong coupling regime. Our results extend the concepts of LEP to quantum thermodynamics. Our study paves the possibility to explore the physical properties of topological QHEs with experiments at a single particle level.

%In the experimental system of a single ion, the components of the topological heat engine are the follows. External thermal baths are controlled by monitoring $\Delta$ and $\gamma_{\mathrm{eff}}$. The minimum and maximum detunings correspond to hot and cold baths, respectively.  The single ion can absorb and release heat by coupling to the effective environments. $\gamma _{\mathrm{eff}}$ can be modified by varying the power of the 854 nm laser, which is tuned to the internal electronic $P_{3/2}-D_{5/2}$ transition. $\Omega$ is adjusted by tuning the power of the 729 nm laser red-tuned to the internal electronic $S_{1/2}-D_{5/2}$ [Fig. \ref{Fig1} (a)].

In our experiment, we execute a QHE cycle consisting of four strokes, as plotted in  Fig. \ref{Fig1}(b). The first stroke which is implemented by increasing the detuning $\Delta$ linearly from its minimum value $\Delta_{min}$ to its maximum value $\Delta_{\rm{max}}$ while $\gamma_\mathrm{eff}$ is kept at its maximum value of $\gamma_{\mathrm{max}}$, is an iso-decay compression from A ($\Delta _{\mathrm{min}},\gamma _{\mathrm{max}}$) to B ($\Delta _{\mathrm{max}},\gamma _{\mathrm{max}}$).  The second stroke is an isochoric heating from B ($\Delta_{\mathrm{max}}, \gamma_{\mathrm{max}}$) to C ($\Delta_{\mathrm{max}},\gamma _{\mathrm{min}}$) and is implemented by decreasing $\gamma_{\mathrm{eff}}$ from $\gamma_{\mathrm{max}}$ to $\gamma_{\mathrm{min}}$, during which the detuning remains at $\Delta _{\mathrm{max}}$.
The third stroke is an iso-decay expansion  from C ($\Delta _{\mathrm{max}},\gamma _{\mathrm{min}}$) to D ($\Delta _{\mathrm{min}},\gamma _{\mathrm{min}}$) implemented by linearly decreasing $\Delta $ from $\Delta _{\mathrm{max}} $ to $\Delta _{\mathrm{min}}$ with $\gamma _{\mathrm{eff}}=\gamma _{\mathrm{min}}$ staying unchanged. Finally, we execute an isochoric cooling from D ($\Delta _{\mathrm{min}}, \gamma _{\mathrm{min}}$) to A ($\Delta_{\mathrm{min}},\gamma_{\mathrm{max}}$) by increasing $\gamma _{\mathrm{eff}}$ from $\gamma _{\mathrm{min}}$ to $\gamma _{\mathrm{max}}$ while keeping $\Delta $ fixed at $\Delta_{\mathrm{min}}$. After the last stroke, we wait for the system reach its steady state and return to its initial state in order to complete a closed cycle.
For this QHE cycle, the 729-nm laser irradiation together with the real environment constitutes the hot and cold baths which correspond to $\Delta_{\mathrm{min}}$ and $\Delta_{\mathrm{max}}$, respectively. Heat exchange between the qubit and the baths is controlled by the detuning $\Delta$ (i.e., controlled by the frequency of the Ti-Sapphire laser). The strength $\Omega$ remains unchanged during the QHE cycle. Thus we may concentrate on the first and third strokes (i.e., the iso-decay processes) for scrutinizing the heat-work exchange of the QHE.

%Quantum working strokes are executed by varying the frequency of the driving laser which helps change the internal energy gap $\Delta$ and the temperature of the working substance. Similarly, quantum isochoric strokes are performed by tuning the effective decay rate $\gamma_\mathrm{eff}$ which controls the heat exchange between the working substance and its thermal baths. As a result, the power of the 854 nm laser helps tune the ion decay, the power and the frequency of the 729 nm laser helps define the four steps of topological operations with four points A ($\Delta _{\mathrm{min}},\gamma _{\mathrm{max}}$), B ($\Delta _{\mathrm{max}},\gamma _{\mathrm{max}}$), C ($\Delta _{\mathrm{max}},\gamma _{\mathrm{min}} $), and D ($\Delta _{\mathrm{min}},\gamma _{\mathrm{min}}$).
The variation of $\gamma _{\mathrm{eff}}$ in the second and fourth strokes (i.e., the isochoric processes) leads to the topological transition between the exact and broken phases. Actually,
an intriguing feature of non-Hermitian systems with exceptional points is the topological structure of the Riemann manifold describing the complex energy of the system,  which leads to state-flip (i.e., state exchange)~\cite{PRL-126-170506} or the accumulation of a geometric phase~\cite{prl-86-787,nature-526-554} when the system parameters are tuned to encircle an EP. We note that this is true for both Hamiltonian exceptional points (HEPs)~\cite{PRL-126-170506} and LEPs~\cite{PRL-127-140504}.  Here, our ability to tune the system parameters to complete a QHE cycle (i.e., a loop in the parameter space) enables us to complete this cycle with or without encircling the LEP (Fig. 1b). The cycle with or without the LEP encircled corresponds to the quantum trajectory traversing the two branches or located in one branch of the Riemann surface (Fig. 1c). As discussed later, this unique feature, along with the dissipative LZS process, leads to higher output work from the engineered QHE.

\begin{figure}[tbph]
\includegraphics[width=8.5 cm]{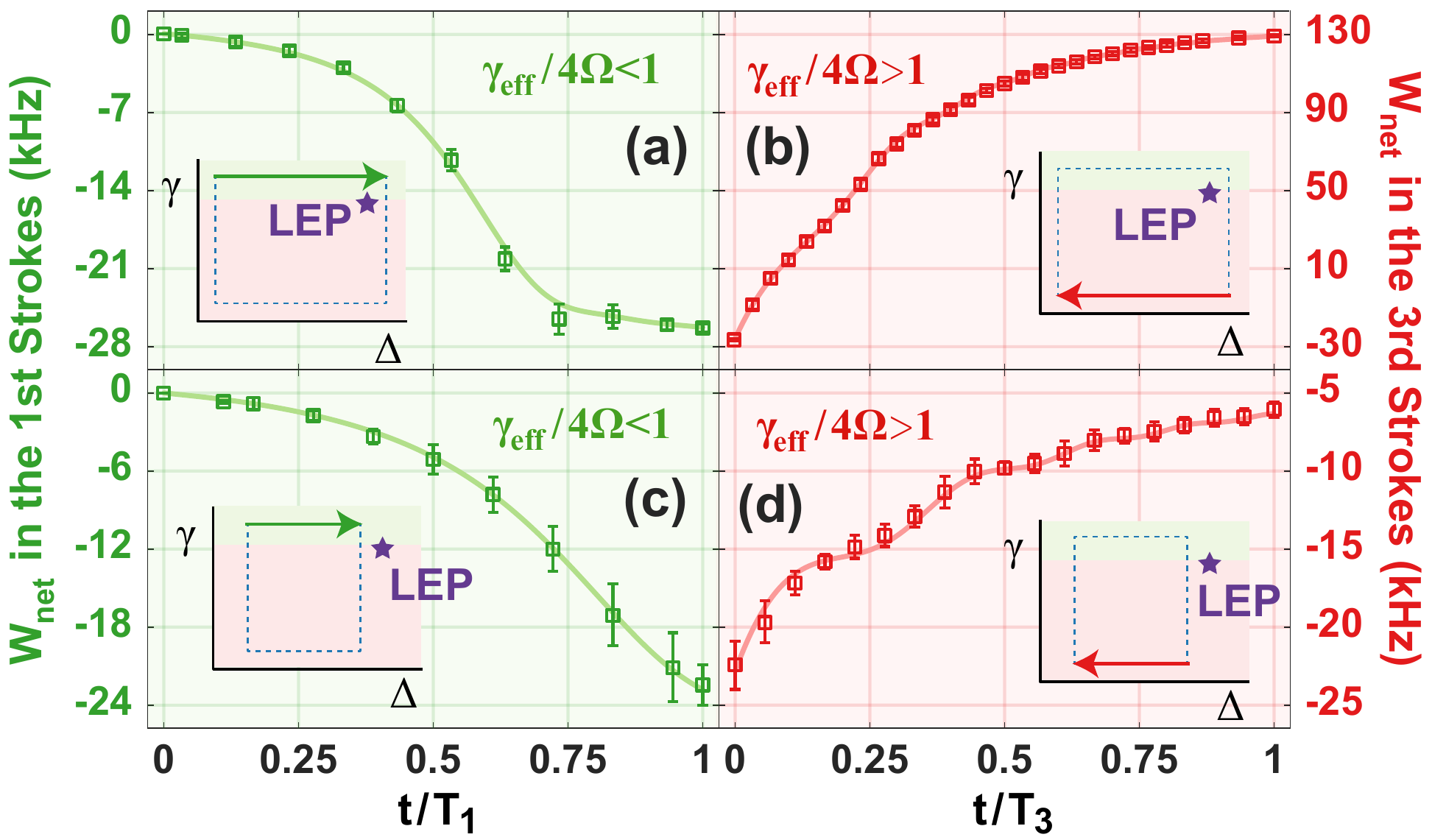}
\caption{ Measured populations of the excited state, i.e., $P_{2}$, in iso-decay strokes, where (a) and (b) are for the first and third strokes, respectively, in the case of encircling the LEP ($\Delta _{\mathrm{min}}/2\pi =-290$ kHz, $\gamma _{\mathrm{max}}=800$ kHz, $\Delta _{\mathrm{max}}/2\pi =10$ kHz, $\gamma_{\mathrm{min}}=130$ kHz). The time is set as $T_{1}=T_{3}=30$ $\mu$s. (c) and (d) represent the first and third strokes, respectively, in the case of not encircling the LEP ($\Delta _{\mathrm{min}}/2\pi =-223$ kHz, $\gamma _{\mathrm{max}}=800$ kHz, $\Delta _{\mathrm{max}}/2\pi =-43$ kHz, $\gamma _{\mathrm{min}}=200$ kHz). The time is $T_{1}=T_{3}=18$ $\mu$s. (e, f, g, h) Net work corresponding to the panels (a, b, c, d), respectively, is defined and explained in the text. In all the panels, $\Omega/2\pi$ is a constant and fixed to be 29 kHz. }
\label{Fig2}
\end{figure}

Experimentally, we execute two QHE cycles as designed in Fig. \ref{Fig1}(b) by elaborately tuning $\gamma_{\rm{eff}}$ using the Ti-sapphire and the semiconductor laser frequencies as two knobs. For convenience of description, we call the red (green) curve with (without) the LEP encircled as a big (small) cycle. To witness the thermodynamic properties, we focus on the iso-decay compression (i.e., the first stroke) and expansion (i.e., the third stroke), in which the heat engine performs work. We first check the big cycle. We observe a hump in the population variation in Fig. \ref{Fig2}(a) and the population oscillation in  Fig. \ref{Fig2}(b). {\color{black}These population variations represent the qubit that absorbs or releases heat from the driven field~\cite{SM}. Therefore, the iso-decay strokes are not adiabatic processes.}
Since the LEP is located at $\Delta=0$, tuning $\Delta$ back and forth through $\Delta=0$ can be understood as an LZS process \cite{LZS}, where the Landau-Zener transition \cite{LZtransition} occurs in the first stroke with detuning tuned from $\Delta_{\mathrm{min}}/2\pi =-290$ kHz to $\Delta_{\mathrm{max}}/2\pi =10$ kHz, and the phase accumulated between transitions, commonly known as the St{\"u}ckelberg phase \cite{Sphase}, may result in constructive or destructive interference in the reverse operation in the third stroke. However, different from the conventional LZS processes that are executed in the absence of decoherence, our QHE execution is subject to decoherence, which modifies the behavior of both the Landau-Zener transition and the St{\"u}ckelberg interference. Thus in the first stroke, the avoided crossing, the typical characteristic of the Landau-Zener transition, is replaced by a crossing between two Riemann sheets, and in the third stroke, we see damped oscillations in the population variation, implying the fringe under decoherence.
In contrast, for the small cycle as witnessed in Fig.~\ref{Fig2}(c,d), the population hump in panel (c) is weak and incomplete due to only a non-resonance transition involved. As a consequence, the population variation in panel (d) is also weak and behaves as the overdamped oscillation.
It should be noted that, for convenience of comparison, we ensure the same rate of the detuning variation in both cycles. As a result, we set the implementation time $T_{1}=T_{3}=30$ $\mu$s for the big cycle, while $T_{1}=T_{3}=18~\mu$s for the small cycle.

The net work is an essential quantity for evaluating the QHE performance. We quantify the net work produced in the two QHE cycles above by defining the net work as $W_{{\rm net}}= -\int\limits_{0}^{t}\rho_{c}(t)dH_{c}$, where $\rho_{c}(t)$ describes the state of the two-level system governed by $H_{c}=\Delta (t)\left\vert 2\right\rangle \left\langle 2\right\vert$ \cite{prl-128-090602}. {\color{black} The corresponding heat of the QHE cycles can be given as $Q_{\mathrm{in(out)}}=\int\limits_{0}^{t}H_{c}(t)d\rho_{c}$ for $d\rho_{c}>0$ ($d\rho_{c}<0$)~\cite{SM}.} Since the isochoric strokes do not perform work, here we only consider the first and third strokes.
Figures.~\ref{Fig2} (e) and (g) illustrate, respectively, the negative net work produced in the first strokes in both the big and small cycles, indicating the fact that the baths perform work on the system in the iso-decay compression, and the Landau-Zener transition in the big cycle leads to more work acquired from the baths. In contrast, the results in Figs.~\ref{Fig2} (f) and (h), corresponding to the third strokes of the big and small cycles, respectively, present the amount of net work accumulated from the starting time of the third stroke counteracting the acquired work in the first stroke. Although the net work in both the big and small cycles increases, it becomes positive only for the big cycle (i.e., net work is still negative at the end of the third stroke for the small cycle). Considering the phase-induced oscillation in Fig.~\ref{Fig2}(b), we attribute the positive net work observed in Fig.~\ref{Fig2}(f) to the coherence accumulated in the St{\"u}ckelberg phase. {\color{black} Note that the corresponding net works of the big and small cycles take the efficiencies $\eta\approx40.65\%$ and $\eta\approx-0.92\%$, respectively.} 

\begin{figure}[tbph]
\includegraphics[width=8.5 cm]{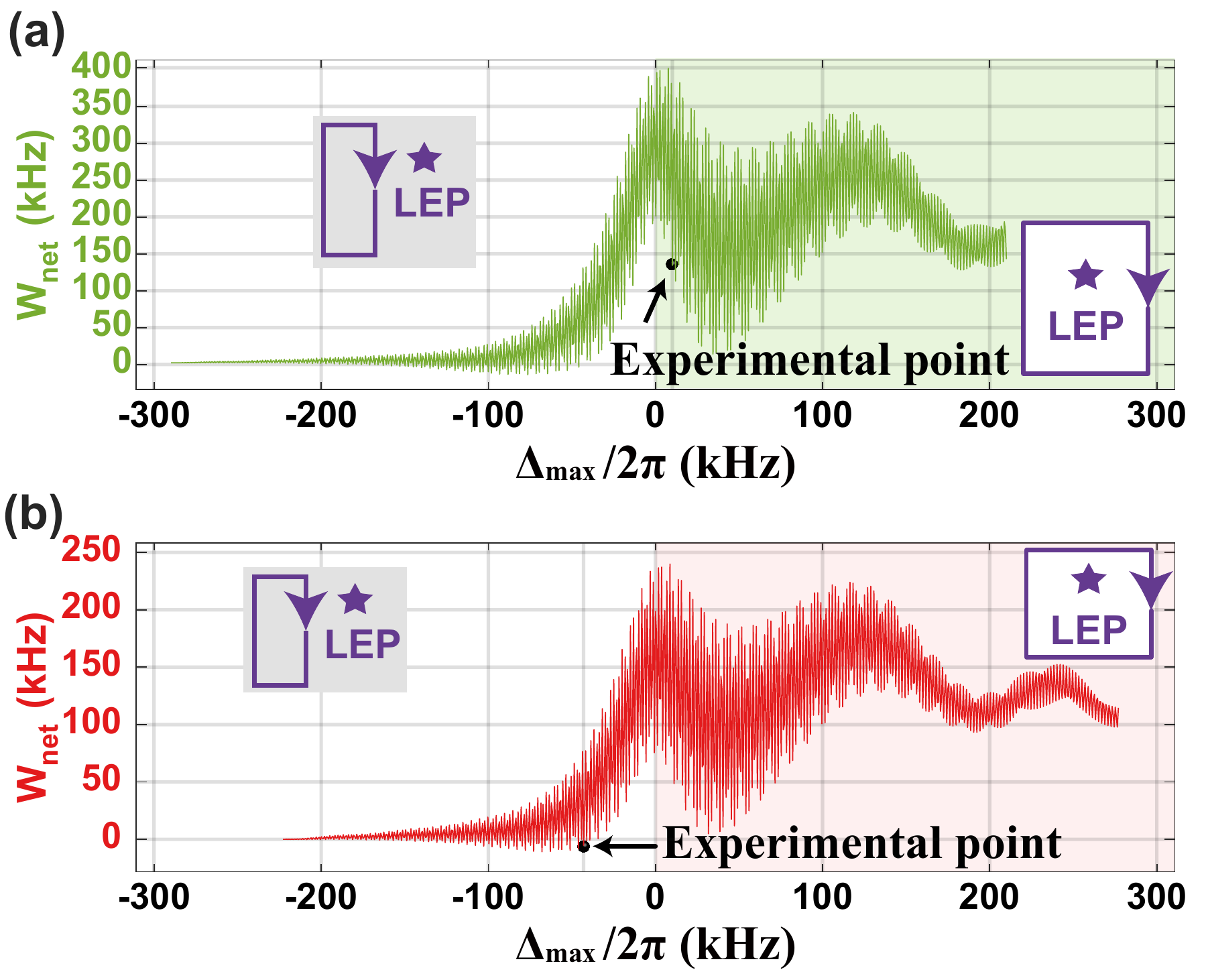}
\caption{Numerical simulation showing the dependence of the net work on $\Delta_{\mathrm{max}}$ of the QHE cycles. (a) The situation covering the big cycle of our experiment, where the QHE cycle starts from the point A with $\Delta _{\mathrm{min}}/2\pi =-290$ kHz and $\gamma_{\mathrm{max}}=800$ kHz. Sweeping the frequency to $\Delta _{\mathrm{max}}/2\pi \leqslant 210$ kHz. So the point C is set with -289.9 kHz $\leqslant\Delta_{\mathrm{max}}/2\pi \leqslant$ 210 kHz and $\gamma_{\mathrm{min}}=$ 130 kHz. (b) The situation involving the small cycle of our experiment, where the QHE cycle starts from A with $\Delta _{\mathrm{min}}/2\pi =-223$ kHz and $\gamma_{\mathrm{max}}=800$ kHz. Sweeping the frequency to $\Delta _{\mathrm{max}}/2\pi \leqslant 277$ kHz. So the point C is set with -222.9 kHz $\leqslant\Delta_{\mathrm{max}}/2\pi \leqslant$ 277 kHz and $\gamma_{\mathrm{min}}=$ 200 kHz. In both the panels, we consider 804 points as the values of the maximal. Black dots indicate the maximal detuning of the two cycles experimentally accomplished. The points A and C referred to can be found in Fig.~\ref{Fig1}, but not labeled here for simplicity. }
\label{Fig3}
\end{figure}

To justify the positive net work observed in Fig.~\ref{Fig2} to be relevant to the LEP, we numerically evaluate the net work of the two QHE cycles in different variations of the detuning in the first and third strokes, checking if $W_{{\rm net}}$ can remain positive once the LEP is enclosed in the QHE cycle. To this end, we fix the point A, i.e., the value of $\Delta_{\rm{min}}$ to be the same as in Fig. \ref{Fig2}, but consider different values of $\Delta_{\rm{max}}$ in the calculation of the first and third strokes.
%where we have assured to have the same ranges of detuning variation in both panels of (a) and (b).
In addition, the effective decays in the second and fourth strokes are varied between 130 kHz and 800 kHz, also the same as in Fig. \ref{Fig2}. Since the maximal values of $\Delta_{{\rm max}}$ considered here are much larger than those considered in the experiments, this numerical result helps understand what happens in our experimental observation. The numerical simulation in Fig. \ref{Fig3} indicates clearly that $W_{\rm{net}}$ can be positive or negative if the LEP is not encircled; while is definitely positive once the LEP is enclosed in the QHE cycle. Moreover, we find in Fig. \ref{Fig3} that the largest net work exists close to the LEP.

Our analysis above neglects the second and fourth strokes which are relevant to the topological transition between the exact and the broken phases induced by the LEP. Although no work is generated during the second and fourth strokes, the topological transitions occurred during the encircling of the LEP in these two strokes, as parts of the QHE cycle, play an important role in the enhancement of the QHE performance. In Fig. \ref{Fig2}, we see the association of the positive net work with the higher population observed in the big cycle. Since lowering $\gamma$ in the second stroke in the big cycle, leading to the topological transition from the broken to the exact phases, helps increasing the population $P_{2}$ (see Fig.~\ref{Fig1}b in comparison with the small cycle),  we attribute the enhanced performance of the QHE to the thermodynamic effects and the topological transition regarding the LEP. We note that previous experimental demonstrations considered topological effects induced by encircling EPs ~\cite{prl-120-146402,prl-123-066405,prl-128-160401}, but not by encircling LEPs. In this paper, we have established a link between the performance of a QHE and the topological effects originating from encircling an LEP. Further exploring such a single-spin system to clarify the LEP-relevant topological effect is highly expected.

In conclusion, we have executed the first QHE by dynamically encircling the LEP in a single trapped ion system. The higher population $P_{2}$, ensuring the positive net work of the QHE, originates from the complete Landau-Zener-St{\"u}ckelberg process and the topological transition across the LEP.
Our experimental observations can be fully understood from the Lindblad master equation, which takes quantum jumps into account, and the LEP represents a proper description of both topological and thermodynamic effects involved in the QHE cycles.  We have also revealed by numerical simulation the maximal net work also relevant to the LEP.
These results would help understand thermodynamic effects in non-Hermitian systems exhibiting exceptional points and the roles of quantum effects in heat-work conversion and working substance-bath interaction in heat engines. Therefore, our study opens interesting possibilities to explore both the thermodynamic and topological effects in the control of the dynamics of the LEP-relevant quantum processes.

This work was supported by Special Project for Research and Development in Key Areas of
Guangdong Province under Grant No. 2020B0303300001, by National Key Research $\&$ Development Program of China under grant No. 2017YFA0304503, by National Natural Science Foundation of China under Grant Nos. U21A20434, 12074390, 11835011, 11734018. S.K.O acknowledges the support from Air Force Office of Scientific Research (AFOSR) Multidisciplinary University Research Initiative (MURI) Award No. FA9550-21-1-0202.  \\

\end{document}